\documentclass[journal,10pt]{IEEEtran}
\usepackage{mathrsfs}
\usepackage{amsfonts}
\usepackage{graphicx,cite,epsfig,amssymb,amsmath}
\usepackage{color,xcolor}
\definecolor{red}{rgb}{1.00, 0.00, 0.00}  
\usepackage{pifont}
\usepackage{stmaryrd}
\usepackage{setspace}
\usepackage{subfigure}
\usepackage{cite}
\usepackage{array}
\usepackage{float}
\usepackage{multirow}
\usepackage{algorithm,algpseudocode}
\usepackage{epstopdf}
\usepackage{mathtools}
\usepackage{diagbox}
\usepackage{booktabs}

\providecommand{\algorithmname}{Algorithm}
\floatname{algorithm}{\protect\algorithmname}
\newcommand{\bm}[1]{\mbox{\boldmath{$#1$}}}
\newcommand{\tabincell}[2]{\begin{tabular}{@{}#1@{}}#2\end{tabular}}

\usepackage{arydshln}
\usepackage{verbatim}
\begin{document}
	
\title{Graph Embedding  based Wireless Link Scheduling with Few Training Samples}
\author{Mengyuan Lee, \IEEEmembership{Graduate Student Member, IEEE,} Guanding Yu, \IEEEmembership{Senior Member, IEEE,} and Geoffrey Ye Li, \IEEEmembership{Fellow, IEEE}
	\thanks{M. Lee and  G. Yu are with the College of Information Science and Electronic Engineering, Zhejiang University, Hangzhou 310027, China. e-mail: \{mengyuan\_lee, yuguanding\}@zju.edu.cn. (\emph{Corresponding author: Guanding Yu})}
	\thanks{G. Y. Li is with the School of ECE, Imperial College London, London, UK. e-mail: geoffrey.li@imperial.ac.uk.}}
\maketitle

\begin{abstract}
Link scheduling in device-to-device (D2D) networks is usually formulated as a non-convex combinatorial problem, which is generally NP-hard and difficult to get the optimal solution. Traditional methods to solve this problem are mainly based on mathematical optimization techniques, where accurate channel state information (CSI), usually obtained through channel estimation and feedback, is needed. To overcome the high computational complexity of the traditional methods and eliminate the costly channel estimation stage, machine leaning (ML) has been introduced recently to address the wireless link scheduling problems. In this paper, we propose a novel graph embedding based method for link scheduling in D2D networks. We first construct a fully-connected directed graph for the D2D network, where each D2D pair is a node  while interference links among D2D pairs are the edges. Then we compute a low-dimensional feature vector for each node in the graph. The graph embedding process is based on the distances of both communication and interference links, therefore without requiring the accurate CSI. By utilizing a multi-layer classifier, a scheduling strategy can be learned in a supervised manner based on the graph embedding results for each node. We also propose an unsupervised manner to train the graph embedding based method to further reinforce the scalability  and develop a  K-nearest neighbor graph representation method to reduce the computational complexity. Extensive simulation demonstrates that the proposed method is near-optimal compared with the existing state-of-art methods but is with only hundreds of training network layouts. It is also competitive in terms of scalability and generalizability to more complicated scenarios.
\end{abstract}



\section{Introduction}
Link scheduling in device-to-device (D2D) networks  is a challenging issue, for which no efficient global optimal algorithm is available yet, especially for the densely deployed network with a large number of mutually interfering links. The goal of link scheduling is to maximize the network utility by activating only a subset of  mutually interfering links at any given time. 

With the help of accurate channel state information (CSI), wireless link scheduling is usually formulated as a non-convex combinatorial optimization problem, which is generally NP-hard and is solved using various mathematical optimization techniques. Some works aim to develop the global optimal algorithms \cite{optimal1,optimal2} but with exponential computational complexity in the worst case. To reduce the computational complexity, sub-optimal algorithms have been developed, including greedy heuristic search algorithm  \cite{flashlinq}, sequential link selection algorithms \cite{itlinq,itlinq2}, iterative  fractional programming algorithm\cite{fplinq}, and heuristic greedy coloring algorithm\cite{graphcolor}. The mathematical optimization methods suffer from three shortcomings. First, the performance of sub-optimal algorithms is hard to control due to the existence of multiple local optima. Second, the computational complexity for both the optimal and sub-optimal methods, such as the iterative algorithms, is too high to meet the real-time requirements. Furthermore, those aforementioned algorithms require accurate CSI, which is usually obtained through channel estimation and feedback. However, for the densely deployed network we consider here, a large number of channels need to be estimated and the channel estimation stage will be both time- and resource-consuming, rendering the difficulty for practical implementation. 

To address the above three issues, we turn to machine learning (ML) techniques for wireless link scheduling. The success of ML in various related fields, such as computer vision and natural language processing, has attracted lots of attention from the wireless communications community recently. ML has already been used in physical layer processing \cite{li1,li2,twostar,add1,add2}, power allocation \cite{shi,mlop2,mlop3,guo}, linear sum assignment problems \cite{wcl}, spectrum sharing \cite{onestar,spc1,spc2,add3},  and user association \cite{user1}. All aforementioned works adopt the widely-used end-to-end learning paradigm.

Meanwhile, another paradigm is to exploit the specific algorithm structures based on ML to simplify resource allocation, for example for cloud radio access networks (Cloud-RANs) \cite{lorm} and D2D communications \cite{minlp}. In this case, resource allocation is still formulated as a mixed integer nonlinear programming (MINLP) problem and imitation learning is used to accelerate the branch-and-bound algorithm, a critical step to solve the MINLP problem.

To avoid CSI requirement in wireless link scheduling, a new ML-based approach, named ``spatial learning", has been developed in \cite{yuwei}. The key idea is first learning the interference pattern of the neighboring transmitters/receivers in the D2D network by using two kernels, and then  learning the optimal scheduling results over multiple feedback stages with the help of the  deep neural networks (DNNs). The whole training process is achieved in an unsupervised manner. The ``spatial learning" is competitive to the sate-of-art FPLinQ algorithm in \cite{fplinq} but needs no accurate CSI. However, the developed approach needs hundreds of thousands of training network layouts, which is difficult to obtain and makes the training process both memory- and time-consuming.

To reduce the number of required training network layouts while maintaining the advantage of requiring no accurate CSI, we propose a new graph embedding based method to deal with the wireless link scheduling problem in D2D networks. Graph embedding is a way to convert the graph data into a low-dimensional space \cite{embedding}. It generates  low-dimensional feature vectors for the whole graph or a part of the graph. Our key idea is to represent the D2D network as a graph and learn the low-dimensional feature vectors for each node  corresponding to a D2D pair. This graph embedding process  learns the interference pattern among different D2D pairs based on the topology of the graph and requires no CSI estimation. Furthermore, the link scheduling problem can be reduced to a binary classification problem since the state of each D2D pair can only be active or inactive. It can be further solved by a multi-layer classifier with low-dimensional feature vectors for each node as input. Parameters of the graph embedding process and the multi-layer classifier are jointly learned  by using the discriminative training method in a supervised manner. Extensive simulation demonstrates that the proposed method can achieve satisfactory performance compared with the FPLinQ algorithm \cite{fplinq} and the ``spatial learning" method \cite{yuwei} but with only hundreds of training network layouts, as summarized in Table \ref{table_com}. In brief, our main contributions are as follows.
\begin{table*}
	\small
	\caption{Comparisons Between Different Methods for Wireless Link Scheduling}
	\label{table_com}
	\centering
	\begin{tabular}{|c|c|c|c|}
		\hline
		Method &\tabincell{c}{ FPLinQ Algorithm\\ \cite{fplinq}} & \tabincell{c}{Spatial Learning \\ \cite{yuwei}} & \tabincell{c}{Graph Embedding\\ Based Method}\\
		\hline
		\hline
		Methodology & \tabincell{c}{Mathematical \\ optimization technique} & ML technique & ML technique\\
		\hline
		Key Idea & Fractional programming & Kernel method \& DNN &Graph embedding\& DNN \\
		\hline
		CSI & Yes & No & No\\
		\hline
		\tabincell{c}{Needed Training\\ Network Layouts }&/ & 	Hundreds of thousands & Hundreds\\
		\hline
		Training Method &/& Unsupervised &Supervised or Unsupervised\\
		\hline
		Scalability  &/& Strong & Good\\
		\hline
		Generalizability  &/& Strong & Good\\
		\hline
		Complexity & $O(L^2)$ & $O(L)$&\tabincell{c}{Fully-connected graph: $O(L^2)$\\ K-nearest neighbor graph: $O(L)$}\\
		\hline
		\tabincell{c}{Generalizability to \\Other Problems}&No &Limited & Good\\
		\hline
	\end{tabular}
\end{table*}

\begin{itemize}
	\item We develop a graph embedding based method for link scheduling in D2D networks. The proposed method requires no accurate CSI and only needs hundreds of training network layouts to achieve the near optimal performance compared with the existing state-of-art methods for link scheduling. As far as we know, this paper is the first attempt to introduce graph embedding method into wireless networks. Our proposed method can be generalized to other problems in wireless networks with appropriate graph representation and feature selection.
	\item We carefully design the node features and edge features for the D2D networks by utilizing the distance information. We further introduce the uniform quantization method to  discretize the continuous distance feature, which reduces the  feature dimension and is essential for implementing graph embedding.  This quantization method can also deal with other continuous features in wireless networks.
	\item To further improve the performance, we adopt the unsupervised learning manner to reinforce the scalability of the graph embedding method. Moreover, the K-nearest neighbor graph representation method is proposed to reduce the computational complexity of the proposed method. 
\end{itemize}

The rest of this paper is organized as follows. In Section II, we introduce the wireless link scheduling problem in D2D networks and formulate it into a non-convex combinatorial problem. In Section III, we develop the graph embedding based method to solve it.  In Section IV, we present test results of the  proposed method. The testing results suggest some shortcomings of the proposed method, which inspire us to make two further improvements in Section V.  Finally, we conclude this paper in Section VI.

\section{Wireless Link Scheduling in D2D Networks}
\begin{figure}[h]
	\centering
	\subfigure[System model.]{
		\begin{minipage}[t]{0.5\linewidth}
			\centering
			\includegraphics[width=2in]{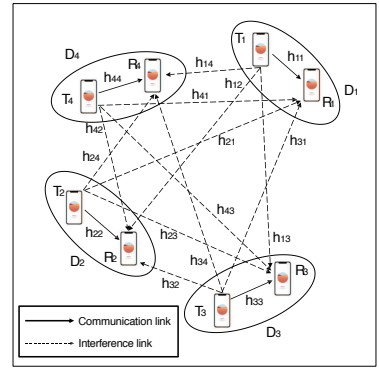}
			\label{fig:sysmodel}
		\end{minipage}%
	}
	\subfigure[Graphical model.]{
		\begin{minipage}[t]{0.5\linewidth}
			\centering
			\includegraphics[width=2in]{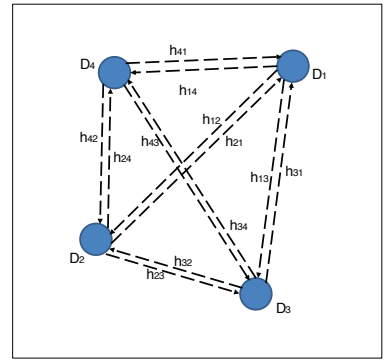}
			\label{fig:graphrepre}
		\end{minipage}
 	}
	\caption{Graph representation process.}
\end{figure}

As depicted in Fig. \ref{fig:sysmodel},  we consider a system with $L$ D2D pairs  in a set $\mathscr{D}=\{D_1, ..., D_L\}$ randomly located in a two-dimensional square region with edge length $d_{\rm area}$. For each D2D pair $D_l \in \mathscr{D}$, we denote its transmitter and receiver as $T_l$ and $R_l$, respectively.  We assume that  each D2D pair $D_l \in \mathscr{D}$ is within a pairwise distance between $d_{\min}$ and $d_{\max}$. We further assume that if $D_l$ is activated, the transmit power of $D_l$ is fixed and denoted as $p_l$. Note that channel multiplexing is not required in this scenario. All D2D pairs reuse the full bandwidth to transmit, which is the same as \cite{fplinq,yuwei}.

As in Fig. \ref{fig:sysmodel}, denote $h_{ll}$ as the channel gain of the communication link of $D_l$, and $h_{lk}$ as the channel gain of the interference link  between $T_l$ and $R_k$. We further introduce $\bm{\rho}= [\rho_{l}]$ as the indicator vector of the D2D pairs' state, with $\rho_{l}=1$ if $D_l$ is activated and $\rho_{l}=0$ otherwise. Then the signal-to-interference-plus-noise ratio (SINR)  of $D_l$ can be written as
$$SINR_{l}(\bm{\rho})=\frac{\rho_{l}p_{l}|h_{ll}|^2}{\sigma_N^2+\sum_{k\neq l} \rho_{k}p_k|h_{kl}|^2},$$
where $\sigma_N^2$ denotes the power of the additive white Gaussian noise (AWGN). Accordingly, the data rate of $D_l$ over bandwidth $B$ can be written as
\begin{equation} \label{rate}
\begin{split}
R_{l}(\bm{\rho}) &= B\log(1+SINR_{l}(\bm{\rho}))\\
&= B \log(1+\frac{\rho_{l}p_{l}|h_{ll}|^2}{\sigma_N^2+\sum_{k\neq l} \rho_{k}p_k|h_{kl}|^2}). 
\end{split}
\end{equation}

From (\ref{rate}), the interference among different D2D pairs will be severe and the data rate will decrease if too many links are activated simultaneously. Therefore, we need to select a subset of links from $\mathscr{D}$ to activate and maximize the throughput of the overall network. If the weighted instantaneous sum rate is used as the objective function, then the wireless link scheduling problem can be formulated as
\begin{eqnarray*}\label{Mresource}
	\hspace{8em}\max_{\bm{\rho}} \sum_{l=1}^L \omega_l R_{l}(\bm{\rho}),\hspace{8.5em}
	\eqref{Mresource_obj} \label{Mresource_obj}
\end{eqnarray*}
\begin{subequations}
	subject to
	\begin{align}
	\rho_{l} \in \{0,1\}, \quad{\forall}l,\label{resource_sub1}
	\end{align}
\end{subequations}
where $\omega_l$ is the weight for the $l$-th D2D pair and can be determined according to fairness or priority in advance.

Problem (\ref{Mresource}) is a combinatorial optimization problem that is difficult to deal with. Traditional approaches \cite{optimal1,optimal2,flashlinq,itlinq,itlinq2,fplinq,graphcolor} to address it are based on various optimization techniques, which require accurate CSI. Recently, ML techniques have been developed to deal with this issue. In \cite{yuwei}, the kernel method has been used to learn the interference pattern among different D2D pairs and perform scheduling with DNNs, which requires no CSI but needs a large amount of training network layouts.

In this paper, we will treat the whole system as a graph. By modeling each D2D pair as a latent variable model, the system is embedded into feature spaces and each D2D pair is represented by a low-dimensional vector. Then the wireless link scheduling will be performed based on the graph embedding results. Our method requires no CSI and can achieve satisfactory performance with much fewer training network layouts compared with the method in  \cite{yuwei}.

\section{Graph Embedding based Wireless Link Scheduling}
In this section, we will develop the graph embedding based method for Problem (\ref{Mresource}). We will first discuss how to model a network into a graph. Then we will introduce how to compute the graph embedding and perform the wireless link scheduling by multi-layer classifier. Finally, we will propose the training process. For simplicity, we assume that $\omega_l=1$ for all D2D pairs in the sequel and will discuss other weight vectors in Section VI.

\subsection{Graph Representation for Wireless Link Scheduling}
We begin by exploring the graph representation method for the network in Fig. \ref{fig:sysmodel}. A weighted graph $G(V,E,\alpha)$ is composed of a set of nodes, $V$, and  a set of edges, $E$. Edge $e(u,v)\in E$ connects two nodes, $u, v \in V$, and has a corresponding weight, $\alpha(u,v)$. If each edge has a direction, the graph is referred to as a directed graph. 

As for the network in Fig. \ref{fig:sysmodel}, each user can be regarded as a node while each link can be regarded as an edge. The weight of each edge could be channel gain, as in Fig. \ref{fig:sysmodel}, or distance between the two nodes of the corresponding link. However, this graph representation method would induce several problems. On the one hand, our aim is to decide whether a D2D pair should be activated. If we regard each user as a node and learn its' corresponding embedding result, we have to combine the embedding results for the transmitter and receiver of a certain D2D pair to make the scheduling decision. How to combine two nodes' embedding results appropriately is tricky. On the other hand, the impact of the communication links and the interference links are totally different in the whole network. However, the graph representation method mentioned above regards them equivalently as an edge in the graph and cannot effectively differentiate between them.

To deal with the aforementioned two problems, we propose to regard each D2D pair as a node and each interference link as an edge to construct the graphical model for the network in Fig. \ref{fig:sysmodel}. The node features and edge weights depend on the channel gains or distances between the two nodes of the corresponding communication/interference links.  Since each edge has a direction, $e(1,2)$ and $e(2,1)$ are two different edges. $e(1,2)$ represents the interference link from $T_1$ to $R_2$ with the direction from node 1 to 2, while $e(2,1)$ represents the interference link from $T_2$ to $R_1$ with the opposite direction. In this way, the original network can be presented as a fully-connected weighted directed graph as illustrated in Fig.  \ref{fig:graphrepre}.

\subsection{Graph Embedding for Wireless Link Scheduling}
After representing the original system into a graph, we will  focus on how to learn graph embedding. Graph embedding is an effective and efficient way to convert the graph data into a low-dimensional space \cite{embedding}. The output of graph embedding is a low-dimensional vector representing the whole graph or a part of the graph, such as a node, an edge, or a substructure. Because our ultimate goal is to decide whether a D2D pair that is represented as a node should be activated, i.e., perform binary classification for each node, the output of graph embedding is the low-dimensional feature vector of each node, which is used for further classification. In this paper, we leverage \emph{structure2vec} in \cite{s2v}, a deep learning architecture over graphs, to achieve the graph embedding process.
\subsubsection{Basic Structure2Vec}
For a \emph{structure2vec} architecture, the output is the $p$-dimensional feature embedding of each node. It achieves this goal by performing nonlinear function mappings iteratively. To be more specific, the \emph{structure2vec} architecture first initializes the feature embedding $\mu_v^{(0)}=0$ for each node in $V$. Then the feature embeddings of all nodes will be updated simultaneously at each iteration by
\begin{eqnarray}
	\mu_v^{(t+1)} = \Gamma(x_v, \{\alpha(u,v)\}_{u\in N(v)},\{{\mu_u^{(t)}}\}_{u\in N(v)}),  \label{update}
\end{eqnarray}
where $x_v$ is the feature of node $v$, $N(v)$ represents the adjacent nodes to node $v$, $\alpha(u,v)$ is the weight of the edge from $u$ to $v$, and $\Gamma$ is a nonlinear function. 

In the updating rule in (\ref{update}), $\{\alpha(u,v)\}_{u\in N(v)}$ and $\{{\mu_u^{(t)}}\}_{u\in N(v)}$  reflect the information of the incoming edges and the neighboring nodes of node $v$, respectively. It is obvious that the feature embedding of each node depends on the specific node feature and the graph topology. Moreover, more update iterations mean that the node features will propagate to more distant nodes. If $T$ iterations are carried out, feature embedding of each node $\mu_v^{(T)}$ will contain the information of its $T$-hop neighborhoods determined by graph topology.

Compared with the widely-used kernel method for feature extraction, the \emph{structure2vec} architecture has the following advantages. First, it uses nonlinear feature mapping instead of the kernel matrix that is tricky to design. Second, the nonlinear function mapping is always small and explicit, which can avoid keeping a huge kernel matrix and can be learned with fewer training network layouts. Finally, the nonlinear function mapping in the \emph{structure2vec} architecture can be learned by stochastic gradient descent, making it efficient to handle extremely large scale datasets. 

\subsubsection{Structure2Vec for Wireless Link Scheduling}
Now we investigate how to leverage the \emph{structure2vec} architecture for wireless link scheduling.

\textit{\romannumeral1. Nonlinear Feature Mapping and Updating Rule:}

To begin with, we choose a specific nonlinear feature mapping. The \emph{structure2vec} architecture can run in a mean field update fashion and is referred to as embedded mean field. For the embedded mean field, we choose the rectified linear unit (ReLU), $\sigma(x) = \max(0,x)$, which can be implemented in the neural network as the nonlinear function mapping. Therefore, the updating rule in (\ref{update}) can be rewritten as
\begin{eqnarray}
\mu_v^{(t+1)} = \sigma(W_1 x_v+ W_2 \sum_{u\in N(v)}\alpha(u,v)+W_3 \sum_{u\in N(v)}\mu_u^{(t)}),  \label{mean}
\end{eqnarray}
where $\bm{W}= \{W_1, W_2, W_3\}$ is the weight set for different information. $\bm{W}$  should be learned with subsequent binary classification task  according to \cite{s2v}. Its training process will be introduced in the following subsection. Once $\bm{W}$ is learned, we can use the pseudo code in Table II for embedded mean field, where the number of iterations, $T$, for the graph embedding computation is tuned via cross validation.  As mentioned above, $T$ iterations mean that the node features can propagate to its $T$-hop neighborhoods. However, the graph for wireless link scheduling is fully-connected, thus each node computes its own feature embedding at the first iteration and propagates its features to other nodes at the second iteration. Therefore, two iterations are usually sufficient for the fully-connected graph to propagate node features over the graph, which coincides with our simulation results in Section IV. {Note that the updating rule in (\ref{mean}) is heuristic and therefore can be further improved in the future work.

From the updating rule in (\ref{mean}), the feature embedding for each D2D pair depends on the communication links between the D2D pair, $x_v$, the interference links to the receiver of the D2D pair,  $\{\alpha(u,v)\}_{u\in N(v)}$, and the feature embeddings of its neighboring D2D pairs, $\{{\mu_u^{(t)}}\}_{u\in N(v)}$.  To be more specific, $x_v$ indicates the communication ability of each D2D pair,  $\{\alpha(u,v)\}_{u\in N(v)}$ represents the interference that each D2D pair receives from its neighbors,  and $\{{\mu_u^{(t)}}\}_{u\in N(v)}$  reflects the interference that each D2D pair causes to its neighbors. Therefore, the feature embedding results of each D2D pair include sufficient information for subsequent link scheduling.

\begin{table}[h]
	\setlength{\abovecaptionskip}{-2pt}
	\setlength{\belowcaptionskip}{-6pt}
	\caption{Embedded Mean Field}
	\begin{algorithm}[H]
		\caption{Embedded Mean Field}
		\label{A1}
		{\normalsize
			\begin{algorithmic}[1]
				\State \textbf{input:} $\bm{W}= \{W_1, W_2, W_3\}$
				\State \textbf{initialization} $\mu_v^{(0)}=0$, for all $v \in V$.
				\For{$t = 1$ to $T$}
				\For{$v \in V$}
				\State $\mu_v^{(t+1)} = \sigma(W_1 x_v+ W_2 \sum_{u\in N(v)}\alpha(u,v)+ \indent \indent \indent \indent W_3 \sum_{u\in N(v)}\mu_u^{(t)})$.
				\EndFor
				\EndFor
				\State \textbf{return} $\{{\mu_v^{(T)}}\}_{v\in V}$.
		\end{algorithmic}}
	\end{algorithm}
\end{table}

\textit{\romannumeral2. Distance Quantization and One-hot Features:}

Another important issue for leveraging the \emph{structure2vec} architecture is to select appropriate node features, $x_v$, and edge weights, $\alpha(u,v)$. As mentioned before, $x_v$ and $\alpha(u,v)$ could be the channel gain or distance between the two nodes of the corresponding communication link as well as interference link, respectively. However, CSI is difficult to estimate in practice. Meanwhile, the optimal wireless link scheduling does not necessarily require the exact CSI and is to a large extent determined by the relative locations of the transmitters and receivers according to \cite{yuwei}.  Therefore, we adopt the distance information of each link as the corresponding features. 

However, directly using the distance between the two nodes of each link as the node and edge features cannot perfectly fit the \emph{structure2vec} architecture. In the \emph{structure2vec} architecture, the dimension of the feature embedding, $\mu_v$, is highly related to that of the node and edge features. The distance of each link lies in  a particular interval but is continuous, so the dimension of distance is infinite. Therefore, we propose to quantify distance to construct discrete features for each node and edge. 

To be more specific, we use $q$ bits to quantify the distance of each link following the uniform quantization method\cite{quan} and construct one-hot features based on the quantization results. We first uniformly divide the quantizer ranges into $2^q$ quantization intervals with indices from $1$ to $2^q$. Then we check the distance of each link. If it lies in the $i$-th interval, the one-hot feature of the corresponding node/edge is a  $2^q$-dimensional vector consisting  of 0s in all cells with the exception of a single 1 in the $i$-th cell. Note that the quantizer range of communication links is $[d_{\min},d_{\max}]$ while that of the interference links is $[0,d_{\rm area}]$. In this way, the dimension of both node and edge features decreases to $2^q$. It is obvious that different numbers of quantization bits will lead to different quantization accuracies and have different influences on the final  scheduling result, as we will discuss later in Section IV.

\subsection{Multi-layer Classifier for Wireless Link Scheduling}
After leveraging the \emph{structure2vec} architecture to compute the graph embedding of the original D2D network, each node in the graph is now represented by a $p$-dimensional vector, which reflects the communication ability and the interference pattern of the corresponding D2D pair. As mentioned above, each node corresponds to a D2D pair and our aim is to decide whether a D2D pair should be activated. This  is a binary classification problem and a link classifier is needed. Note that the \emph{structure2vec} architecture is implemented as a neural network according to \cite{s2v}. Given that the parameters of the graph embedding and the link classifier should be learned jointly, we use a multi-layer classifier, i.e., a DNN,  to solve the binary classification problem. The input layer of the classifier consists of $p$ neurons and  takes the node embedding feature, $\mu_v^{(T)}$, as input. The output layer consists of $2$ neurons, indicating the probability of activation or not, respectively.

\subsection{Training Process}
The overall network includes two parts: the graph embedding and the multi-layer classification.  Accordingly, the parameters to be learned also consist of two parts: the embedding parameters, $\bm{W}= \{W_1, W_2, W_3\}$, of the algorithm in Table II and the parameters for the classifier, $F$.  In this paper, we adopt the   discriminative training method in \cite{s2v} to learn these two parts of parameters jointly.

The key idea of the discriminative training is to learn $\bm{W}$ and $F$ jointly for the ultimate task in a supervised manner. Suppose that we have a training dataset $\mathscr{T} = \{x_n,y_n\}_{n=1}^N$, where $x_n$ is the graphical model for the D2D network and  $y_n$ is the corresponding scheduling result, respectively. The training dataset is generated by running the state-of-art FPLinQ algorithm proposed in \cite{fplinq} with a  maximum iteration of 100\footnote{Although FPLinQ is sub-optimal, experiment results in \cite{yuwei} suggest that the scheduling outputs of the FPLinQ algorithm after 100 iterations show good numerical performance. Therefore, using FPLinQ algorithm to generate training dataset is acceptable.}. To be more specific, $y_n$ is a $L$-dimensional vector and $y_n \in \{0,1\}^L$, where $y_n(l) = 1$ if $D_l$ is activated, and $y_n(l) = 0$, otherwise. With the feature embedding procedure proposed in Table II, each graph $x_n$ is represented as a set of embedding vectors $ \{\mu_v^n\}_{v \in V}$. Our goal is to learn a classifier, $F$, mapping $ \{\mu_v^n\}_{v \in V}$ to  $y_n$. We denote $z_l^n$ as the one-hot representation of $y_n(l)$, i.e., $z_l^n = (1,0)$ if $y_n(l) = 0$, and $z_l^n= (0,1)$ if $y_n(l) = 1$. We further denote $F(\mu_v^n)$ as $\tilde{z}_l^n$.  By adopting cross entropy as the loss function, the joint optimization for the embedding parameters and the classifier parameters can be written as
\begin{eqnarray}
 \min_{\textbf{W},F} \sum_{n=1}^{N}\sum_{l=1}^{L}-z_l^n(0)\log \tilde{z}_l^n(0)-z_l^n(1)\log \tilde{z}_l^n(1). \label{loss}
\end{eqnarray}
By optimizing the objective function in (\ref{loss}) using any optimization algorithm for the neural networks, $\bm{W}$ and $F$ can be learned jointly in an end-to-end paradigm.

\section{Performance Test Results}
In this section, we will test the performance of the proposed graph embedding based method for  wireless link scheduling. We utilize existing open-source code for the  \emph{structure2vec} architecture\footnote{https://github.com/Hanjun-Dai/pytorch\_structure2vec/tree/master/s2v\_lib} and all other codes are implemented in python 3.6 except the FPLinQ algorithm that is implemented in Matlab.\footnote{The simulation source code is available at https://github.com/mengyuan-lee/graph\_embedding\_link\_scheduling.} 

\subsection{Simulation Setup}
We consider a 500 m by 500 m two-dimensional square area with 50 D2D pairs as depicted in Fig. \ref{fig:sysmodel}. According to \cite{yuwei}, the transmitter of each D2D pair is distributed uniformly in the area and the corresponding receiver is distributed in a disk centered by the transmitter with uniform pairwise distance between 2 m and 65 m. We use the short-range outdoor model ITU-1411 with a distance-dependent path-loss \cite{channel} as the channel model. Our system parameters are consistent with those in \cite{yuwei} and are summarized in Table \ref{table2}, where the carrier frequency and the antenna height are used in  ITU-1411 model to calculate the path-loss. 
\begin{table}
	\normalsize
	\caption{System Parameters}
	\label{table2}
	\centering
	\begin{tabular}{|c|c|}
		\hline
		Parameter & Value  \\
		\hline
		\hline
		Edge length, $d_{\rm area}$ & 500 m \\
		\hline
		D2D distance, $d_{\min}$, $d_{\max}$ & 2 m, 65 m \\
		\hline
		Noise spectral density & -169 dBm/Hz\\
		\hline
		Bandwidth, $B$ & 5 MHz\\
		\hline
		Carrier frequency & 2.4 GHz\\
		\hline
		Antenna height & 1.5 m\\
		\hline
		Transmit power of activation link, $p_l$ & 40 dBm\\
		\hline
	\end{tabular}
\end{table}

We use a three-layer neural network as the binary classifier.  Meanwhile, we use 500 network layouts for training and 1,000 network layouts for testing. Note that 500 training network layouts in the scenario with 50 links actually include 25,000 training data points. We tune the hidden layer size in $\{16,32,64\}$, the embedding dimension, $p$, in $\{16,32,64\}$ and the number of iterations, $T$, for the algorithm in Table II in $\{1,2,3,4,5\}$  by hold-out validation and report the average performance over 1,000 testing network layouts\footnote{We have also tried to use 5,000 network layouts for the following tests as \cite{yuwei}. The testing results are very close to those by using 1,000 testing network layouts, which suggests that 1,000 testing network layouts can lead to convincing results.}.  Note that the number of iterations, $T$, should be set as 2 according to  the analysis in Section III-B. The performance of the proposed method with different $T$  is summarized in Table \ref{table_T}, where \emph{classifier accuracy} and \emph{average sum rate} are two valuation metrics for the proposed method. Specifically, \emph{classifier accuracy} reflects the similarity between the scheduling pattern output by the proposed method and the FPLinQ algorithm, and \emph{average sum rate} is the normalized sum rate achieved by the proposed method with respect to that achieved by the FPLinQ algorithm. From Table \ref{table_T}, $T\geq 2$ achieves higher sum rate than $T=1$. Furthermore, when $T\geq 2$, a larger $T$ would not always lead to better performance but would bring about higher complexity. Therefore, we select $T=2$ in the following tests.

\begin{table}  
\small
\caption{Performance with Different Numbers of Iterations} 
\label{table_T} 
\centering  
\begin{tabular}{|c|c|c|c|c|c|}  
\hline
\tabincell{c}{Number of\\ Iterations, $T$}  & 1 &2 &3 &4 &5\\
\hline
\tabincell{c}{Classifier \\Accuracy}& 0.8123 &0.8101 &0.8140 &0.8110 &0.8150\\
\hline
\tabincell{c}{Average \\Sum Rate} & 0.9462 &0.9521 &0.9507&0.9501 &0.9532\\
\hline
\end{tabular}  
\end{table} 

We also use batch normalization \cite{bn} to avoid vanishing gradient problem and early stopping \cite{early_stop} to avoid overfitting on small training dataset. Moreover, we set the hidden layer size as 64, the embedding dimension as 32,  and the quantization bits of  features, $q=3$ for the test of this section. Note that, $q$ and $p$ are different. They correspond to the input and the output dimensions of the graph embedding process, respectively. Meanwhile, we adopt the adaptive moment estimation (Adam) algorithm as the optimization algorithm \cite{adam}. Our neural network parameters are summarized in Table \ref{table}. 

\begin{table}
	\normalsize
	\caption{Neural Network Parameters}
	\label{table}
	\centering
	\begin{tabular}{|c|c|}
		\hline
		Parameter & Value  \\
		\hline
		\hline
		Hidden layer size & 64 \\
		\hline
		Embedding dimension, $p$ & 32\\
		\hline
		Number of iterations, $T$ &2\\
		\hline
		Number of training network layouts & 500\\
		\hline
		Number of testing network layouts& 1,000\\
		\hline
		Quantization bits, $q$ & 3 \\
		\hline
	\end{tabular}
\end{table}

\subsection{Convergence Performance}
To evaluate the convergence performance of the proposed method, we plot the \emph{classifier accuracy} and the \emph{average sum rate} for the training process of the scenario with 50 D2D pairs. The results are shown in Fig. \ref{fig:mis}.

From Fig. \ref{fig:mis}, \emph{average sum rate} and \emph{classifier accuracy} both converge after only 30-40 training epochs with 500 training network layouts. These results suggest that the convergence speed of the proposed method is fast and the training stage is not time-consuming, which is preferred in practice. Furthermore, from the figure, higher \emph{classifier accuracy} mostly leads to higher \emph{average sum rate} but with some mismatch, that is, sometimes higher \emph{classifier accuracy} may result in lower \emph{average sum rate} such as the area circled in Fig. \ref{fig:mis}.  This kind of mismatch is reasonable but is not preferred in the supervised training process. Our immediate goal of the training process is to increase the accuracy even if our ultimate goal is to maximize the average sum rate. We hope that better training accuracy can definitely lead to higher average sum rate. We will  deal with this mismatch issue in Section V-A to produce better results.
\begin{figure}[h]
	\centering
	\includegraphics[width=0.8\linewidth, height=0.22\textheight]{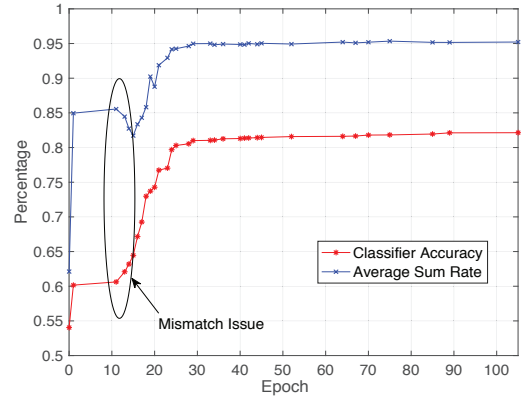}
	\caption{Convergence performance for the scenarios with 50 D2D pairs using 500 training network layouts.}
	\label{fig:mis}
\end{figure}

\subsection{Impact of the Number of Training Network Layouts}
We investigate how many training network layouts are needed to learn a good network. In practice, labeled training network layouts are sometimes difficult and expensive to obtain. Therefore, methods requiring a small training dataset are preferred in practice, especially in wireless communication systems where a large number of real data is hard to get. On the other hand, requiring fewer training network layouts generally implies faster training speed and less memory consumption, which is also preferred in real-time implementation. 

We test the performance of the proposed method using different numbers of training network layouts on the scenario with 50 D2D pairs. We set the number of quantization bits, $q=3$, and report the average performance on 1,000 testing network layouts. The results are summarized in Table \ref{table3}. As shown in Table \ref{table3}, the performance of the proposed method fluctuates with the number of training network layouts. As we can image, using more training network layouts will not always lead to better performance and hundreds of training network layouts are enough.

\begin{table}  
	\small
	\caption{Performance with Different Numbers of Training Network Layouts} 
	\label{table3} 
	\centering  
	\begin{tabular}{|c|c|c|c|c|c|}  
		\hline
		\tabincell{c}{Number of \\Training\\ Network\\ Layouts}  & 200 &500 &1,000 &1,500 &2,000\\
		\hline
		\tabincell{c}{Classifier \\Accuracy}& 0.7975 &0.8101 &0.8103 &0.8044 &0.8050\\
		\hline
		\tabincell{c}{Average \\Sum Rate} & 0.9336 & 0.9521 &0.9498 &0.9512 &0.9456\\
		\hline
	\end{tabular}  
\end{table}

\subsection{Comparison with Existing Methods for Link Scheduling}
We compare it with the FPLinQ algorithm \cite{fplinq}, the ``spatial learning" method  \cite{yuwei}, and several benchmark algorithms used in \cite{yuwei}. 

The detailed comparative results are summarized in Table \ref{compare_all}, where we use 500 training network layouts for our proposed method and 1,000 testing network layouts for all presented methods.   From Table \ref{compare_all}, our method can achieve 95.21\% of the sum rate produced by FPLinQ algorithm by using only 500 training network layouts, without explicitly knowing the CSI. It can outperform the \emph{strongest link first} algorithm, the \emph{random active} algorithm, and the \emph{all active} algorithm. Compared with the ``spatial learning" method in \cite{yuwei} that uses 800,000 training network layouts to achieve 98.36\% of the  sum rate produced by FPLinQ algorithm, we use far fewer training network layouts but with only 3.15\% loss of sum rate. Compared with the greedy algorithm where accurate CSI is needed to re-evaluate the sum rate every time a new link is activated or deactivated, our proposed method is much more explicit and has only 1.87\% loss of sum rate even without the accurate CSI. 

\begin{table}  
	\small
	\caption{Comparison of Different Algorithms for Link Scheduling} 
	\label{compare_all} 
	\centering  
	\begin{tabular}{|c|c|c|c|c|c|c|c|}  
		\hline
		Method &CSI &Average Sum Rate\\
		\hline
		\tabincell{c}{ Graph Embedding \\Based Method} &  No& 0.9521\\
		\hline
		\tabincell{c}{FPLinQ  \cite{fplinq}} & Yes & 1.0000\\
		\hline
		\tabincell{c}{ Spatial Learning \cite{yuwei}} &No &0.9836 \\
		\hline
		Greedy &Yes&0.9708\\
		\hline
		\tabincell{c}{Strongest Link First} &Yes&0.8203 \\
		\hline
		\tabincell{c}{Random Active}&No&0.4747 \\
		\hline
		\tabincell{c}{All Active}&No&0.5418\\
		\hline
	\end{tabular}  
\end{table}

\subsection{Scalability to Scenarios with Different Topologies}
We study the scalability of the proposed method to scenarios with different topologies, especially to more complicated scenarios. There are many existing ML methods  for  wireless communication systems whose performance deteriorates sharply with the complexity of the problems. In this subsection, we first test how the proposed method performs when the numbers of D2D pairs change and then demonstrate its performance on scenarios with different pairwise distances.
\subsubsection{Scalability to Scenarios with Different Numbers of D2D Pairs}
The number of D2D pairs is highly related to the complexity of wireless link scheduling. More D2D pairs would lead to more complicated interference, which results in the difficulty for scheduling.  We  test the performance of the proposed method on scenarios with different numbers of D2D pairs.  For each scenario, we generate 500 training network layouts, set the number of quantization bits, $q=3$, and report the average performance on 1,000 testing network layouts. The results are summarized in Table \ref{table4}.  
\begin{table}  
	\footnotesize
	\caption{Performance on Scenarios with Different Numbers of D2D Pairs} 
	\label{table4} 
	\centering  
	\begin{tabular}{|c|c|c|c|c|c|c|c|}  
		\hline
		\tabincell{c}{Number of \\D2D Pairs} & 10 &30 &50 &80 &100 &500  \\
		\hline
		\tabincell{c}{Classifier \\Accuracy}& 0.8792 &0.8092 &0.8101 &0.8203 &0.8260 & 0.8895  \\
		\hline
		\tabincell{c}{Average \\Sum Rate} & 0.9751 & 0.9608 &0.9521 &0.9308 &0.9226 & 0.8658  \\
		\hline
	\end{tabular} 
\end{table} 

From Table \ref{table4}, the \emph{average sum rate} only decreases by 2.95\% when doubling the number of D2D pairs from 50 to 100.  Moreover, the \emph{average sum rate} is still acceptable for the scenario with 500 links, whose link number is 10 times larger than the scenario with 50 links. Apparently, our proposed method performs well for scenarios with different numbers of D2D pairs with only 500 training network layouts. Note that the classifier accuracy with 500 D2D pairs is higher than the others. This may come from the fact that most links should be deactivated with the scenario with a large number of D2D pairs. It does not mean a better performance on the average sum rate either.  Furthermore, we apply the hyperparameters we selected on the scenario with 50 D2D pairs to every case presented in Table \ref{table4}. The results are satisfactory, suggesting that it is feasible  to select hyperparameters based on small networks and then apply to large networks in practice.

\subsubsection{Scalability to  Scenarios with Different Pairwise Distances} 
As mentioned in \cite{yuwei}, the pairwise distance has significant effect on the achievable rate. Wireless link scheduling  tends to activate short links, therefore the distribution of pairwise distances has significant influence on the scheduling performance. We test the performance of the proposed method on scenarios with different pairwise distances and the results are summarized in Table \ref{table5}. We still generate 500 training network layouts, set the number of quantization bits, $q=3$, and report the average performance on 1,000 testing network layouts for each scenario.

\begin{table}  
	\small
	\caption{Performance  on  Scenarios with Different Pairwise Distances}
	\label{table5} 
	\centering  
	\begin{tabular}{|c|c|c|c|c|}  
		\hline
		\tabincell{c}{Pairwise Distance \\$d_{\min} \sim d_{\max}$(m)}& 2 $\sim$ 65 &10 $\sim$ 50& 30 $\sim$ 70  &all 30 \\
		\hline
		Classifier Accuracy& 0.8101 &0.7548 &0.7321 &0.6994\\
		\hline
		Average Sum Rate & 0.9521 & 0.9225 &0.9090 &0.8041\\
		\hline
	\end{tabular}  
\end{table} 

From Table \ref{table5},  the performance of the proposed method deteriorates with the decrease of the pairwise distribution interval and it can be explained as follows. The distribution of pairwise distances directly influences the diversity of node features. Bigger pairwise distribution interval means larger diversity in node features. If the pairwise distance is the same for each D2D pair as in the cases where all the pairwise distances are 30 m, there are no node features but only edge features, causing it more difficult to learn wireless link scheduling. We will solve this issue in Section V-A to improve the scalability of the proposed method.

\subsection{Impact of the Number of Quantization Bits}
In the proposed graph embedding based method, distance quantization is important and crucial. As mentioned in Section III-B, the number of quantization bits directly influences the quantization accuracies and the ultimate scheduling results. In this subsection, we will test the influence of the number of quantization bits. We generate 500 training network layouts and report the average performance on 1,000 testing network layouts for the scenarios with 50 D2D pairs. The results are summarized in Table \ref{table6}.
\begin{table}  
	\small
	\caption{Performance with Different Numbers of Quantization Bits}
	\label{table6} 
	\centering  
	\begin{tabular}{|c|c|c|c|c|c|}  
		\hline
		\tabincell{c}{Number of \\Quantization \\Bits}& 2&3 & 4 & 5& 6\\
		\hline
		\tabincell{c}{Classifier \\Accuracy}& 0.7765&0.8101 &0.8223 &0.8113& 0.8127\\
		\hline
		\tabincell{c}{Average \\Sum Rate} & 0.9401 &0.9521 &0.9522 &0.9500& 0.9502\\
		\hline
	\end{tabular}   
\end{table} 

As shown in Table \ref{table6}, both \emph{classifier accuracy} and \emph{average sum rate} first increase at a decreasing rate and then fluctuate with the increase of the number of quantization bits. The results suggest that increasing the number of quantization bits under certain threshold can improve the performance of the proposed method. However, more quantization bits over the threshold value will not lead to further improvement. This result can be explained as follows. The quantization accuracy certainly increases while increasing the number of quantization bits. Then the quantized features suffer from less distortion and include more information for learning optimal scheduling.  Therefore, the performance improves but the model gets more complicated due to the increase of features' dimension. When increasing the number of quantization bits to certain threshold value, the information included by the quantized feature is exactly sufficient to learn the optimal scheduling. Further increase  on the number of quantization bits would lead to information redundancy and cannot further improve the wireless link scheduling performance.

Inspired from the above observation, we can get some insights about implementation in practice. There exists a trade-off between model complexity and performance. Under certain threshold, using more quantization bits leads to better performance but more complicated models while fewer quantization bits lead to the opposite results. Therefore, we need to choose an appropriate number of quantization bits in practice according to our specific goals instead of simply using as many quantization bits as possible.

\subsection{Generalizability to Scenarios with Different Topologies}
Generalizability is another important property of ML techniques. It is different from the scalability we have discussed above. Scalability focuses on the performance of our proposed method on more complicated scenarios. To test the scalability, a new model needs to be trained for each testing scenario with the training network layouts whose topologies are the same as the testing scenario. However, generalizability focuses on how a trained model performs on unknown scenarios. To test the generalizability, we do not need to train a new model for each testing scenario but just applying a trained model from a certain scenario whose topology can be different from the testing scenarios. Both of them are important for ML techniques while good generalizability is generally more difficult to meet.

Our proposed method can be easily generalized to scenarios with different topologies only if the number of quantization bits is fixed. We fix the number of quantization bits, $q = 3$, and train  the model with 500 samples from the scenario with 50 D2D pairs. We then apply the model on the  scenarios with different numbers of D2D pairs. The results are summarized in Table \ref{table7}. In the table, \emph{full training} means using the model trained with 500 samples whose topology is the same as testing network layouts, and \emph{generalization} means using the model trained with 500 samples from the scenario with 50 D2D pairs. As shown in Table \ref{table7},  there only exist 2.24\% and 3.42\% loss of the average sum rates for the scenarios with 10 and 100 D2D pairs, respectively. The results suggest that our proposed method has a good generalizability to both scenarios with smaller and larger topologies.

\begin{table}  
	\small
	\caption{Average Sum Rate on Scenarios with Different Numbers of D2D Pairs Using Different Modes } 
	\label{table7} 
	\centering 
		\begin{tabular}{|c|c|c|c|c|c|}  
			\hline
			\tabincell{c}{Number of \\D2D Pairs} & 10 &30 &50 &80 &100 \\
			\hline
			Full Training &0.9751 & 0.9608 &0.9521 &0.9308 &0.9226 \\
			\hline
			Generalization& 0.9527 & 0.9622 & 0.9521 &0.8972 &0.8884 \\
			\hline
	\end{tabular}  
\end{table}

\subsection{Impact of the Shadowing}
As mentioned above, all aforementioned testings adopt the ITU-1411 outdoor channel model where only distance-dependent path-loss is considered. In this part, we add shadowing into the channel model and test the performance of the proposed method with different values of shadowing standard deviation. 

We still fix the number of quantization bits, $q = 3$, and set the number of D2D pairs as 50.  We report the average performance over 1,000 testing network layouts and the \emph{average sum rate} results are summarized in Table \ref{table_shadow}. In the table, \emph{full training} means using the model trained with 500 network layouts whose channels are with the same shadowing as the testing network layouts, and \emph{generalization} means using the model trained with 500 network layouts  whose channels are without shadowing. From the table, the average sum rate of the proposed method decreases with the increase of shadowing standard deviation, which is intuitive. However, the average sum rates achieved by \emph{full training}  and \emph{generalization} are very close to each other. The performance gain by using \emph{full training} is less than 1\% for all presented case. These results can be explained as follows. On the one hand, our proposed method only uses the distance information to do link scheduling. Therefore, shadowing is a stochastic variable to the proposed method that has not been learned. When the shadowing standard deviation increases, the performance of the proposed method will definitely deteriorate. On the other hand, the training dataset $\mathscr{T}$ is composed of $\{x_n,y_n\}_{n=1}^N$, where $x_n$ is the graphical model for the D2D network and  $y_n$ is the corresponding scheduling result from the FPLinQ algorithm, respectively. When we add shadowing information to the training network layouts, $y_n$ will change but $x_n$ will remain the same. Therefore, the shadowing information is only indirectly included in the labels of the training dataset. It may help reinforce the performance but the gain will not be too much. Therefore, including the shadowing information in the training network layouts is not helpful. The \emph{full training} method has the similar performance to the \emph{generalization} method.
\begin{table} 
	\small
	\caption{Performance  with Different Shadowing Standard Deviations } 
	\label{table_shadow} 
	\centering 
		\begin{tabular}{|c|c|c|c|c|c|}  
			\hline
			\tabincell{c}{Shadowing \\Standard\\ Deviation (dB)}&0 & 3 &5 &8 &10 \\
			\hline
			Full Training  &0.9488 & 0.9280 & 0.8848  &0.8155 &0.7644\\
			\hline
			Generalization &0.9446 & 0.9259& 0.8742& 0.8145& 0.7572\\
			\hline
	\end{tabular}  
\vspace{-1.5em}
\end{table} 

The similar performance loss also exists for the ``spatial learning" algorithm in \cite{yuwei}, whose performance drops significantly when Rayleigh fast fading is added to the testing channels. This is a universal shortcoming for the link scheduling method without explicit CSI. From the above  results, we may need to include the shadowing information in both $x_n$ and $y_n$ to overcome this shortcoming. In this way, the node and edge features should not only be the distances between nodes of the corresponding communication/interference links. Note that, CSIs of the communication links are much easier to estimate than those of the interference links in practice. We may include the CSIs of the communication links into the node features to deal with the performance loss.  It is a very interesting and important issue for future work.

\subsection{Computational Complexity Analysis}
In this subsection, we analyze the computational complexity for the graph embedding based method and compare it with the FPLinQ algorithm in \cite{fplinq} and the ``spatial learning" in \cite{yuwei}. 

We consider a scenario with $L$ D2D pairs. For the FPLinQ algorithm, the dominant computation for each iteration is the matrix multiplication with the  $L \times L$ channel coefficient matrix. Assuming that the number of iterations is fixed, the total computational complexity for it is $O(L^2)$. For the kernel based ``spatial learning" in \cite{yuwei}, the computational complexity is $O(L)$ under the scenarios of fixed region size. Our proposed method includes two steps: computing graph embedding and performing binary classification. Both steps perform nonlinear function mapping. Note that there are two loops for the graph embedding algorithm in Table II. If the number of iterations, $T$, is fixed, the computational complexity for graph embedding computation is $O(L^2)$.  Meanwhile, the computational complexity for classification is $O(L)$ if the structure of the multi-layer classifier is fixed. Therefore, the computational complexity for our proposed method is $O(L^2)$.  

Based on the aforementioned analysis, our proposed method has the similar computational complexity with FPLinQ algorithm but does not need explicit CSI that is hard to obtain in practice. On the other hand, our proposed method is not competitive in terms of computational complexity compared to the ``spatial learning" in \cite{yuwei}, but it needs far fewer training network layouts to achieve satisfactory performance as shown in Section IV-D.  We will further discuss this problem and reduce the computational complexity of the proposed method in Section V-B.

\section{Discussion and Further Improvement}
Based on the test results and analysis in Section IV, we come up with the following three important questions:
\begin{itemize}
	\item[i)] how to choose an appropriate training goal to match the ultimate goal,
	\item[ii)] how to strengthen the scalability of the proposed method to  scenarios with different pairwise distances, and
	\item[iii)] how to reduce the computational complexity of the proposed method. 
\end{itemize}  
In this section, we will propose two improvements to address these issues.

\subsection{Graph Embedding based Wireless Link Scheduling in the Unsupervised Manner}
We focus on the first two problems mentioned above in this subsection. The simplest method to address the first problem is to use the sum rate in (\ref{Mresource})  directly as the loss function and maximize it in an unsupervised manner. Meanwhile, as mentioned in \cite{yuwei}, unsupervised learning is competitive for the scenarios with small pairwise distance distribution intervals. It can also avoid using sub-optimal FPLinQ algorithm to generate training dataset and obtain better performance. Therefore, we will figure out how to train the graph embedding based method in the unsupervised manner and then test the corresponding performance.

\subsubsection{Implementing Graph Embedding based Method in the Unsupervised Manner}
To develop the graph embedding based method in the unsupervised manner,  the graph representation, graph embedding, and multi-layer classifier are the same as what we have introduced in Section III while the training process needs to be modified. 

We still make use of the discriminative training method.  Suppose that the training dataset for unsupervised learning is $\mathscr{T}_u = \{x_n\}_{n=1}^N$, where $x_n$ is the graphical model for the D2D network and  no corresponding scheduling result is included. By running the feature embedding procedure proposed in Table II, each graph $x_n$ is represented as a set of embedding vectors $ \{\mu_v^n\}_{v \in V}$. Our goal is still to learn the classifier, $F$, and we  denote $F(\mu_v^n)$ as $\tilde{z}_l^n$.  To realize the unsupervised learning, the new optimization problem for embedding parameters and classifier parameters can be formulated as
\begin{gather}
\min_{\textbf{W},F} \sum_{n=1}^{N} ( \frac{1}{ \sum_{l=1}^{L} B \log(1+\frac{\tilde{z}_l^n(1)p_{l}|h_{ll}|^2}{\sigma_N^2+\sum_{k\neq l} \tilde{z}_k^n(1)p_k|h_{kl}|^2})} \notag \\ 
\hspace{13em} - \omega_ {\rm loss} \sum_{l=1}^{L} \log \tilde{z}_l^n(0)). \label{loss2}
\end{gather}
The first part in (\ref{loss2}) is the reciprocal of the objective function in Problem (\ref{Mresource}). By minimizing this part, we actually maximize the sum rate. Note that the inverse of the  sum rate can be also used as the loss function \cite{yuwei}. We use the reciprocal of the sum rate in this paper because negative loss functions are less often encountered than positive ones in the machine learning community. The second part of (\ref{loss2}) is the penalty term for full activation case, where $ \omega_ {\rm loss}$ is its weight and can be tuned in the training process. When the  sum rate is large, the gradient of the loss function in (\ref{loss2}) is very small and may get stuck at some local optima. By observation, all D2D pairs are likely to be activated at the same time while using the unsupervised learning method especially for scenarios with the same D2D pairwise distance. Since there is a lack of node features when the scenarios contain links of the same distances, the learning process will be confused to converge to the local optima where all the links are activated.  Therefore, we use the penalty term in (\ref{loss2}) to avoid getting stuck at the full activation local optimum. The full activation problem  less likely happens in the supervised learning method since the labels in the training dataset for supervised learning can offer some additional node features even if all D2D pairs are of the same link distance. Furthermore, $ \omega_ {\rm loss}$ should be carefully selected. We will first set it to be 0 and tune it from $\{0.005,0.01,0.02\}$ if there exists the full activation issue.

We optimize the objective function in (\ref{loss2}). In this way, $\bm{W}$ and $F$ can be learned together in an unsupervised manner.

\subsubsection{Performance Test}
In this part, we do some tests for the graph embedding based method in an unsupervised manner. Specifically, we pay attention to the scalability of the unsupervised learning method. As in Section IV-E, we test how the unsupervised learning method performs on scenarios with different numbers of D2D pairs and with different pairwise distances, respectively. We generate 500 training network layouts, set the number of quantization bits, $q=3$, and report the average performance on 1,000 testing network layouts for each scenario. The parameters for wireless system and graph embedding based network are the same as what are summarized in Tables \ref{table2} and \ref{table}. The results are summarized in Tables \ref{table8} and \ref{table9}.  The $ \omega_ {\rm loss}$ is set to be 0.005 for the scenarios where all the pairwise distances are 30 m and is set to be 0 for all other cases presented.

As shown in Table \ref{table8}, the average sum rate achieved by the unsupervised learning method is almost the same as that achieved by the supervised learning method. The performance differences are less than 0.7\% for all the presented cases. The results suggest that the unsupervised learning method is not competitive in terms of the scalability to scenarios with different numbers of D2D pairs. However, from Table \ref{table9}, it generally achieves higher sum rate than the supervised learning method in terms of the scalability to scenarios with different pairwise distances. Specifically, the performance gain increases with the decrease of the pairwise distribution interval. In the scenario where all the pairwise distances are 30 m, the average sum rate increases by 7.06\% using the unsupervised learning method. Therefore, implementing the graph embedding based method in the unsupervised manner is an effective method to strengthen the scalability of the proposed method to  scenarios with different pairwise distances.
\begin{table}  
	\footnotesize
	\caption{Performance  with  Different Numbers of D2D Pairs Using Different Training Manners} 
	\label{table8} 
	\centering  
	\begin{tabular}{|c|c|c|c|c|c|c|c|}  
		\hline
		\tabincell{c}{Number of\\ D2D Pairs} & 10 &30 &50 &80 &100 &500 \\
		\hline
		\tabincell{c}{Supervised \\Learning}  &  0.9751 & 0.9608 &0.9521 &0.9308 &0.9226 &0.8658\\
		\hline
		\tabincell{c}{Unsupervised\\ Learning} &0.9739 &0.9628 &0.9534 &0.9371 &0.9281 &0.8645 \\
		\hline
	\end{tabular} 
\end{table} 

\begin{table}  
	\small
	\caption{ Performance with Different Pairwise Distances Using Different Training Manners}
	\label{table9} 
	\centering 
	\begin{tabular}{|c|c|c|c|c|}  
		\hline
		\tabincell{c}{Pairwise Distance \\$d_{\min} \sim d_{\max}$(m)}& 2 $\sim$ 65 &10 $\sim$ 50& 30 $\sim$ 70  &all 30 \\
		\hline
		Supervised Learning&  0.9521 & 0.9225 &0.9090 &0.8041\\
		\hline
		Unsupervised Learning & 0.9534 &0.9310 &0.9241 &0.8747\\
		\hline
	\end{tabular} 
\end{table} 

\subsubsection{Comparison between Supervised and Unsupervised Learning Method}
As mentioned before, the unsupervised learning method can strengthen the scalability  to scenarios with different pairwise distances  of the graph embedding based method. The intuitive explanations compromise two aspects. On the one hand, the FPLinQ algorithm's results are not optimal, which may limit the  performance of supervised learning. On the other hand, different scheduling patterns may lead to the same sum rate but different classifier accuracies. Therefore, directly optimizing the sum rate in an unsupervised manner is preferred rather than optimizing the accuracy based on the FPLinQ algorithm's results.  However, the convergence speed of the unsupervised learning method is slow and it takes more time to train the network. Furthermore, the supervised learning method can avoid the full activation problem. In a nutshell, both methods have their own advantages and disadvantages. We need to carefully select appropriate learning method in practice.

\subsection{K-Nearest Neighbor Graph for Graph Embedding based Wireless Link Scheduling}
In this subsection, we focus on how to reduce the computational complexity of the proposed method. As mentioned in Section IV-I, the computational complexity of our proposed method depends on two procedures: the computation of graph embedding and the classification. Decreasing the computational complexity of the graph embedding computation process is the key point, which is mainly affected by the graph structure. Therefore, we propose to use the K-nearest neighbor graph to replace the fully-connected graph to solve this problem.

\subsubsection{K-Nearest Neighbor Graph Representation Method}
We modify the graph representation process proposed in Section III-A. We construct a K-nearest neighbor graph instead of the fully-connected graph for the D2D system depicted in Fig. \ref{fig:sysmodel}, which means we only consider the top K nearest transmitters to a certain D2D pair's receiver. The K-nearest neighbor graph representation method is reasonable since the interference caused by a transmitter on the considered D2D pair can be negligible if the transmitter is too far away from the considered D2D pair's receiver. In this way, the number of neighbors for each node is fixed and the computational complexity for graph embedding computation process in Table II decreases to $O(L)$. Therefore, the computational complexity of the proposed method also decreases to $O(L)$, which is the same as that of the ``spatial learning" method in \cite{yuwei}.  

\subsubsection{Performance Test}
We do some tests for the graph embedding based method by using the K-nearest neighbor graph representation method. The parameters for wireless system and graph embedding network are the same as what are summarized in Tables \ref{table2} and \ref{table}, and we adopt the supervised training method for the following test.

\textit{\romannumeral1. Impact of the values of K}

We first test the influence of different K values. We  set the  number of quantization bits, $q=3$, and generate 500 training network layouts and 1,000 testing network layouts on the scenarios with 50 D2D pairs. The results are summarized in Table \ref{table11}. From the table, the \emph{classifier accuracy} for all cases fluctuates around 81\% and the \emph{average sum rate} for all cases fluctuates around 95\%. The results suggest that using K-nearest neighbor graph representation method will not definitely induce worse performances compared with the fully-connected graph representation method since the dominant interference information is maintained. Therefore, the K-nearest neighbor graph representation method is effective to reduce the computational complexity of the proposed graph embedding approach without loss of performance.
\begin{table}  
	\footnotesize
	\caption{Performance of K-Nearest Neighbor Graph  \protect\\ Representation Method with Different  $K$}
	\label{table11} 
	\centering  
	\begin{tabular}{|c|c|c|c|c|c|}  
		\hline
		\tabincell{c}{Value of \\$K$}& 10 &20&30&40 & \tabincell{c}{49\\(Fully-connected)} \\
		\hline
		\tabincell{c}{Classifier \\Accuracy}&0.8138& 0.8146&0.8097&0.8099  &0.8101\\
		\hline
		\tabincell{c}{Average\\ Sum Rate} &0.9529 &0.9524&0.9484&0.9503&0.9521\\
		\hline
	\end{tabular}  
\end{table} 

\textit{\romannumeral2. Scalability Test}

In this part, we focus on the scalability of the K-nearest neighbor graph representation method. First, we test how it performs on the scenarios with different numbers of D2D pairs and the results are summarized in Table \ref{table12}. We set the  number of quantization bits, $q=3$, generate 500 training network layouts, and report the average performance on 1,000 testing network layouts for each presented scenarios. We compare the performance of the 10-nearest neighbor graph representation method and the fully-connected graph representation method.  From Table \ref{table12}, the average sum rates achieved by  both graph representation methods are very close to each other for the scenarios with 30 and 50 links. However, the 10-nearest neighbor graph representation method can achieve a slightly better performance for the scenarios with more links. In particular, the average sum rate achieved by the 10-nearest neighbor graph representation method increases by 4.55\% for the scenario with 500 D2D pairs, as compared to the fully-connected representation method. The results suggest that the performance gain achieved by the 10-nearest neighbor graph representation method increases for scenarios with more D2D pairs. The process of constructing K-nearest neighbor graphs discards some negligible information and simplifies the network to be learned.  Therefore, with the same number of training network layouts, the 10-nearest neighbor graph representation method performs better than the fully-connected graph representation method, especially for scenarios with more D2D pairs.
\begin{table}  
	\footnotesize
	\caption{Performance of Different Graph Representation Methods on\protect\\  Scenarios with Different  Numbers of D2D Pairs} 
	\label{table12} 
	\centering  
	\begin{tabular}{|c|c|c|c|c|c|c|c|}  
		\hline
		\tabincell{c}{Number of \\D2D Pairs} & 10 &30 &50 &80 &100 &500 \\
		\hline
		\tabincell{c}{Fully-\\Connected\\ Graph}  & 0.9751 & 0.9608 &0.9521 &0.9308 &0.9226  &0.8658 \\
		\hline
		\tabincell{c}{10-Nearest\\ Neighbor \\Graph} &/ &0.9614 &0.9529 &0.9409 &0.9353 &0.9113 \\
		\hline
	\end{tabular} 
\vspace{-1em} 
\end{table}

We also test how the  K-nearest neighbor graph representation method performs on the scenarios with different D2D pairwise distances,  and the results are shown in Table \ref{table13}. We still set the  number of quantization bits, $q=3$, generate 500 training network layouts, and report the average performance on 1,000 testing network layouts for each presented scenario.  We again compare the performance of the 10-nearest neighbor graph representation method and the fully-connected graph representation method. As shown in Table \ref{table13}, the 10-nearest neighbor graph representation method performs better than the fully-connected method in terms of the scalability to scenarios with different pairwise distances, especially for the scenario where all the pairwise distances are 30 m. This result seems not intuitive but is reasonable. As mentioned in Section IV-E, the distribution of pairwise distances directly impacts the diversity of node features, and smaller pairwise distribution interval makes it more difficult to learn wireless link scheduling due to the lack of diversity for each node. However, the K-nearest neighbor graph representation method can reinforce the diversity of different nodes because each node is now connected to different K nodes instead of being fully connected to all other nodes. In other words, the K-nearest neighbor graph is asymmetric and its nodes  have more diversity, which makes it perform better on the scenarios with different D2D pairwise distances. 
\begin{table}  
	\small
	\caption{Performance of Different Graph Representation Methods \protect\\ on Scenarios with Different  Pairwise Distances }
	\label{table13} 
	\centering 
	\begin{tabular}{|c|c|c|c|c|}  
		\hline
		\tabincell{c}{Pairwise Distance \\$d_{\min} \sim d_{\max}$(m)}& 2 $\sim$ 65 &10 $\sim$ 50& 30 $\sim$ 70  &all 30 \\
		\hline
		\tabincell{c}{Fully-Connected \\Graph} &  0.9521 & 0.9225 &0.9090 &0.8041\\
		\hline
		\tabincell{c}{10-Nearest\\ Neighbor Graph}& 0.9529 &0.9308 & 0.9176& 0.8718\\
		\hline
	\end{tabular}   
\end{table} 

\section{Conclusions and Future Directions}
This paper proposes an ML technique to deal with wireless link scheduling in D2D networks. The key idea is using graph embedding to extract the interference pattern for each D2D pair to perform scheduling. By representing the D2D network as a fully-connected directed graph where each node corresponds to a specific D2D pair and each edge corresponds to the interference link, we can compute the low-dimensional feature vector for each node by graph embedding. Then, a multi-layer classifier can be used to perform scheduling to near optimum based on the embedding results of each node. The proposed method only exploits the relative locations of D2D pairs and avoids the costly channel estimation process. Extensive experiment results have demonstrated that the proposed method only needs hundreds of training network layouts and has satisfactory scalability and generalizability to the scenarios with different numbers of D2D pairs. However, it only has limited scalability to scenarios with different pairwise distances. To further improve the scalability, we have also proposed to train the graph embedding based method in the unsupervised manner. Moreover, we have introduced the K-nearest neighbor graph representation method to enable the proposed method to run in linear computational complexity, which is preferred in practical implementation.

However, the current work still represents a preliminary step towards making use of the graph embedding method for wireless link scheduling problems. First, we have assumed that all  D2D pairs are of the same weight. While the weights are not equal for each D2D pair, our proposed method needs to be modified. However, simply incorporating the weights as a new feature does not work well as presented in other existing works \cite{shi,yuwei}. Therefore, attempting to solve proportional fairness scheduling by making use of the graph embedding method is an very interesting issue for fine-tuning our proposal. Meanwhile, the performance of the proposed method drops significantly when shadowing is introduced to the testing channels.  It is possible to reinforce the proposed method by using partial CSIs. However, our proposed uniform quantization method is not suitable for CSI. Therefore, how to utilize CSIs of the communication links to reinforce the performance of our proposed method is another important future direction.  Moreover, the updating rule for the graph embedding process is heuristic. Modifying the update rules may lead to better performance. Also,  the proposed method is not as scalable and generalizable as the ``spatial learning" method, how to find other effective methods and even adopting new ML techniques to deal with this issue are very important future directions.

To put it into nutshell, this paper suggests that graph embedding is potential for optimization tasks in wireless networks, especially when the optimization results mainly depend on the topology of the networks. With carefully designed graph representation method and selected features, the graph embedding outputs combined with DNNs and other ML methods, such as reinforcement learning, can achieve satisfactory performance compared with the state-of-art methods.

\end{document}